# In-Datacenter Performance Analysis of a Tensor Processing Unit™

Norman P. Jouppi, Cliff Young, Nishant Patil, David Patterson, Gaurav Agrawal, Raminder Bajwa, Sarah Bates, Suresh Bhatia, Nan Boden, Al Borchers, Rick Boyle, Pierre-luc Cantin, Clifford Chao, Chris Clark, Jeremy Coriell, Mike Daley, Matt Dau, Jeffrey Dean, Ben Gelb, Tara Vazir Ghaemmaghami, Rajendra Gottipati, William Gulland, Robert Hagmann, C. Richard Ho, Doug Hogberg, John Hu, Robert Hundt, Dan Hurt, Julian Ibarz, Aaron Jaffey, Alek Jaworski, Alexander Kaplan, Harshit Khaitan, Daniel Killebrew, Andy Koch, Naveen Kumar, Steve Lacy, James Laudon, James Law, Diemthu Le, Chris Leary, Zhuyuan Liu, Kyle Lucke, Alan Lundin, Gordon MacKean, Adriana Maggiore, Maire Mahony, Kieran Miller, Rahul Nagarajan, Ravi Narayanaswami, Ray Ni, Kathy Nix, Thomas Norrie, Mark Omernick, Narayana Penukonda, Andy Phelps, Jonathan Ross, Matt Ross, Amir Salek, Emad Samadiani, Chris Severn, Gregory Sizikov, Matthew Snelham, Jed Souter, Dan Steinberg, Andy Swing, Mercedes Tan, Gregory Thorson, Bo Tian, Horia Toma, Erick Tuttle, Vijay Vasudevan, Richard Walter, Walter Wang, Eric Wilcox, and Doe Hyun Yoon
*Google, Inc., Mountain View, CA USA*
Email: {`jouppi,cliffy,nishantpatil,davidpatterson`} @google.com

**Abstract**
**Many architects believe that major improvements in cost-energy-performance must now come from domain-specific hardware. This paper evaluates a custom ASIC—called a *Tensor Processing Unit (TPU)*— deployed in datacenters since 2015 that accelerates the inference phase of neural networks (NN). The heart of the TPU is a 65,536 8-bit MAC matrix multiply unit that offers a peak throughput of 92 TeraOps/second (TOPS) and a large (28 MiB) software-managed on-chip memory. The TPU's deterministic execution model is a better match to the 99th-percentile response-time requirement of our NN applications than are the time-varying optimizations of CPUs and GPUs (caches, out-of-order execution, multithreading, multiprocessing, prefetching, …) that help average throughput more than guaranteed latency. The lack of such features helps explain why, despite having myriad MACs and a big memory, the TPU is relatively small and low power. We compare the TPU to a server-class Intel Haswell CPU and an Nvidia K80 GPU, which are contemporaries deployed in the same datacenters. Our workload, written in the high-level TensorFlow framework, uses production NN applications (MLPs, CNNs, and LSTMs) that represent 95% of our datacenters' NN inference demand. Despite low utilization for some applications, the TPU is on average about 15X - 30X faster than its contemporary GPU or CPU, with TOPS/Watt about 30X - 80X higher. Moreover, using the GPU's GDDR5 memory in the TPU would triple achieved TOPS and raise TOPS/Watt to nearly 70X the GPU and 200X the CPU.**

## 1. Introduction to Neural Networks

The synergy between the large data sets in the cloud and the numerous computers that power it has enabled a renaissance in machine learning. In particular, *deep neural networks* (DNNs) have led to breakthroughs such as reducing word error rates in speech recognition by 30% over traditional approaches, which was the biggest gain in 20 years [Dea16]; cutting the error rate in an image recognition competition since 2011 from 26% to 3.5% [Kri12] [Sze15] [He16]; and beating a human champion at Go [Sil16]. Unlike some hardware targets, DNNs are applicable to a wide range of problems, so we can reuse a DNN-specific ASIC for solutions in speech, vision, language, translation, search ranking, and many more.

Neural networks (NN) target brain-like functionality and are based on a simple artificial neuron: a nonlinear function (such as `max(0, value)`) of a weighted sum of the inputs. These artificial neurons are collected into layers, with the outputs of one layer becoming the inputs of the next one in the sequence. The "deep" part of DNN comes from going beyond a few layers, as the large data sets in the cloud allowed more accurate models to be built by using extra and larger layers to capture higher levels of patterns or concepts, and GPUs provided enough computing to develop them.

The two phases of NN are called *training* (or learning) and *inference* (or prediction), and they refer to development versus production. The developer chooses the number of layers and the type of NN, and training determines the weights. Virtually all training today is in floating point, which is one reason GPUs have been so popular. A step called *quantization* transforms floating-point numbers into narrow integers—often just 8 bits—which are usually good enough for inference. Eight-bit integer multiplies can be 6X less energy and 6X less area than IEEE 754 16-bit floating-point multiplies, and the advantage for integer addition is 13X in energy and 38X in area [Dal16].

Three kinds of NNs are popular today:
1. *Multi-Layer Perceptrons* (MLP): Each new layer is a set of nonlinear functions of weighted sum of all outputs (*fully connected*) from a prior one, which reuses the weights.
2. *Convolutional Neural Networks* (CNN): Each ensuing layer is a set of of nonlinear functions of weighted sums of spatially nearby subsets of outputs from the prior layer, which also reuses the weights.
3. *Recurrent Neural Networks* (RNN): Each subsequent layer is a collection of nonlinear functions of weighted sums of outputs and the previous state. The most popular RNN is *Long Short-Term Memory* (LSTM). The art of the LSTM is in deciding what to forget and what to pass on as state to the next layer. The weights are reused across time steps.

Table 1 shows two examples of each of the three types of NNs—which represent 95% of NN inference workload in our datacenters—that we use as benchmarks. Typically written in TensorFlow [Aba16], they are surprisingly short: just 100 to 1500 lines of code. Our benchmarks are small pieces of larger applications that run on the host server, which can be thousands to millions of lines of C++ code. The applications are typically user-facing, which leads to rigid response-time limits.

Each model needs between 5M and 100M weights (9th column of Table 1), which can take a lot of time and energy to access. To amortize the access costs, the same weights are reused across a *batch* of independent examples during inference or training, which improves performance.

This paper describes and measures the *Tensor Processing Unit* (*TPU*) and compares its performance and power for inference to its contemporary CPUs and GPUs. Here is a preview of the highlights:
- Inference apps usually emphasize response-time over throughput since they are often user-facing.
- Due to latency limits, the K80 GPU is underutilized for inference, and is just a little faster than a Haswell CPU.
- Despite having a much smaller and lower power chip, the TPU has 25 times as many MACs and 3.5 times as much on-chip memory as the K80 GPU.
- The TPU is about 15X - 30X faster at inference than the K80 GPU and the Haswell CPU.
- Four of the six NN apps are memory-bandwidth limited on the TPU; if the TPU were revised to have the same memory system as the K80 GPU, it would be about 30X - 50X faster than the GPU and CPU.
- The performance/Watt of the TPU is 30X - 80X that of contemporary products; the revised TPU with K80 memory would be 70X - 200X better.
- While most architects have been accelerating CNNs, they represent just 5% of our datacenter workload.

| Name | LOC | Layers | | | | | Nonlinear function | Weights | TPU Ops / Weight Byte | TPU Batch Size | % of Deployed TPUs in July 2016 |
|---|---|---|---|---|---|---|---|---|---|---|---|
| | | FC | Conv | Vector | Pool | Total | | | | | |
| MLP0 | 100 | 5 | | | | 5 | ReLU | 20M | 200 | 200 | 61% |
| MLP1 | 1000 | 4 | | | | 4 | ReLU | 5M | 168 | 168 | |
| LSTM0 | 1000 | 24 | | 34 | | 58 | sigmoid, tanh | 52M | 64 | 64 | 29% |
| LSTM1 | 1500 | 37 | | 19 | | 56 | sigmoid, tanh | 34M | 96 | 96 | |
| CNN0 | 1000 | | 16 | | | 16 | ReLU | 8M | 2888 | 8 | 5% |
| CNN1 | 1000 | 4 | 72 | | 13 | 89 | ReLU | 100M | 1750 | 32 | |

**Table 1.** Six NN applications (two per NN type) that represent 95% of the TPU's workload. The columns are the NN name; the number of lines of code; the types and number of layers in the NN (FC is fully connected, Conv is convolution, Vector is self-explanatory, Pool is pooling, which does nonlinear downsizing on the TPU; and TPU application popularity in July 2016. One DNN is RankBrain [Cla15]; one LSTM is a subset of GNM Translate [Wu16]; one CNN is Inception; and the other CNN is DeepMind AlphaGo [Sil16][Jou15].

## 2. TPU Origin, Architecture, and Implementation

Starting as early as 2006, we discussed deploying GPUs, FPGAs, or custom ASICs in our datacenters. We concluded that the few applications that could run on special hardware could be done virtually for free using the excess capacity of our large datacenters, and it's hard to improve on free. The conversation changed in 2013 when a projection where people use voice search for 3 minutes a day using speech recognition DNNs would require our datacenters to double to meet computation demands, which would be very expensive to satisfy with conventional CPUs. Thus, we started a high-priority project to quickly produce a custom ASIC for inference (and bought off-the-shelf GPUs for training). The goal was to improve cost-performance by 10X over GPUs. Given this mandate, the TPU was designed, verified [Ste15], built, and deployed in datacenters in just 15 months. (Space limits the amount and the level of detail on the TPU in this paper; see [Ros15a], [Ros15b], [Ros15c], [Ros15d], [Tho15], and [You15] for more.)

Rather than be tightly integrated with a CPU, to reduce the chances of delaying deployment, the TPU was designed to be a coprocessor on the PCIe I/O bus, allowing it to plug into existing servers just as a GPU does. Moreover, to simplify hardware design and debugging, the host server sends TPU instructions for it to execute rather than fetching them itself.

Hence, the TPU is closer in spirit to an FPU (floating-point unit) coprocessor than it is to a GPU.

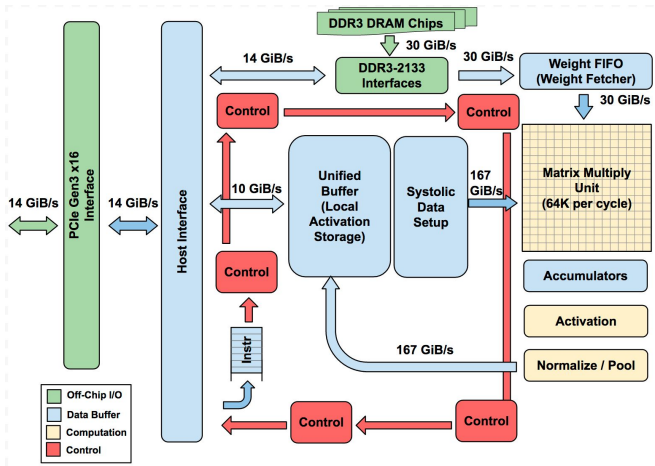 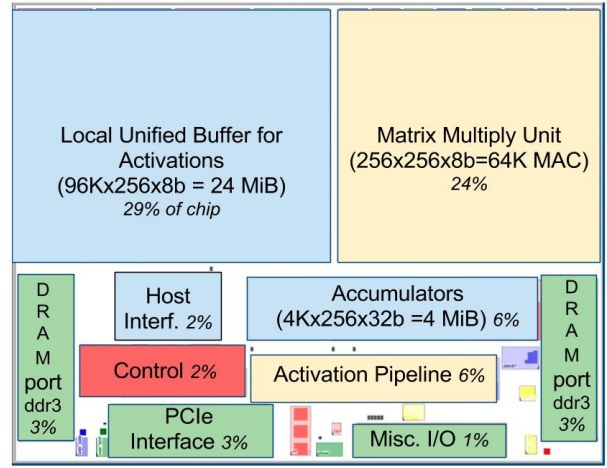

**Figure 1.** TPU Block Diagram. The main computation part is the yellow Matrix Multiply unit in the upper right hand corner. Its inputs are the blue Weight FIFO and the blue Unified Buffer (UB) and its output is the blue Accumulators (Acc). The yellow Activation Unit performs the nonlinear functions on the Acc, which go to the UB.

**Figure 2.** Floor Plan of TPU die. The shading follows Figure 1. The light (blue) data buffers are 37% of the die, the light (yellow) compute is 30%, the medium (green) I/O is 10%, and the dark (red) control is just 2%. Control is much larger (and much more difficult to design) in a CPU or GPU

The goal was to run whole inference models in the TPU to reduce interactions with the host CPU and to be flexible enough to match the NN needs of 2015 and beyond, instead of just what was required for 2013 NNs. Figure 1 shows the block diagram of the TPU.

The TPU instructions are sent from the host over the PCIe Gen3 x16 bus into an instruction buffer. The internal blocks are typically connected together by 256-*byte*-wide paths. Starting in the upper-right corner, the *Matrix Multiply Unit* is the heart of the TPU. It contains 256x256 MACs that can perform 8-bit multiply-and-adds on signed or unsigned integers. The 16-bit products are collected in the 4 MiB of 32-bit *Accumulators* below the matrix unit. The 4MiB represents 4096, 256-element, 32-bit accumulators. The matrix unit produces one 256-element partial sum per clock cycle. We picked 4096 by first noting that the operations per byte need to reach peak performance (roofline knee in Section 4) is ~1350, so we rounded that up to 2048 and then duplicated it so that the compiler could use double buffering while running at peak performance.

When using a mix of 8-bit weights and 16-bit activations (or vice versa), the Matrix Unit computes at half-speed, and it computes at a quarter-speed when both are 16 bits. It reads and writes 256 values per clock cycle and can perform either a matrix multiply or a convolution. The matrix unit holds one 64 KiB tile of weights plus one for double-buffering (to hide the 256 cycles it takes to shift a tile in). This unit is designed for dense matrices. Sparse architectural support was omitted for time-to-deploy reasons. Sparsity will have high priority in future designs.

The weights for the matrix unit are staged through an on-chip *Weight FIFO* that reads from an off-chip 8 GiB DRAM called *Weight Memory* (for inference, weights are read-only; 8 GiB supports many simultaneously active models). The weight FIFO is four tiles deep. The intermediate results are held in the 24 MiB on-chip *Unified Buffer*, which can serve as inputs to the Matrix Unit. A programmable DMA controller transfers data to or from CPU Host memory and the Unified Buffer.

Figure 2 shows the floor plan of the TPU die. The 24 MiB Unified Buffer is almost a third of the die and the Matrix Multiply Unit is a quarter, so the datapath is nearly two-thirds of the die. The 24 MiB size was picked in part to match the pitch of the Matrix Unit on the die and, given the short development schedule, in part to simplify the compiler (see Section 7). Control is just 2%. Figure 3 shows the TPU on its printed circuit card, which inserts into existing servers like an SATA disk.

As instructions are sent over the relatively slow PCIe bus, TPU instructions follow the CISC tradition, including a repeat field. The average clock cycles per instruction (CPI) of these CISC instructions is typically 10 to 20. It has about a dozen instructions overall, but these five are the key ones:
1. `Read_Host_Memory` reads data from the CPU host memory into the Unified Buffer (UB).
2. `Read_Weights` reads weights from Weight Memory into the Weight FIFO as input to the Matrix Unit.
3. `MatrixMultiply/Convolve` causes the Matrix Unit to perform a matrix multiply or a convolution from the Unified Buffer into the Accumulators. A matrix operation takes a variable-sized B*256 input, multiplies it by a 256x256 constant weight input, and produces a B*256 output, taking B pipelined cycles to complete.

4. `Activate` performs the nonlinear function of the artificial neuron, with options for ReLU, Sigmoid, and so on. Its inputs are the Accumulators, and its output is the Unified Buffer. It can also perform the pooling operations needed for convolutions using the dedicated hardware on the die, as it is connected to nonlinear function logic.
5. `Write_Host_Memory` writes data from the Unified Buffer into the CPU host memory.

The other instructions are alternate host memory read/write, set configuration, two versions of synchronization, interrupt host, debug-tag, nop, and halt. The CISC MatrixMultiply instruction is 12 bytes, of which 3 are Unified Buffer address; 2 are accumulator address; 4 are length (sometimes 2 dimensions for convolutions); and the rest are opcode and flags.

The philosophy of the TPU microarchitecture is to keep the matrix unit busy. It uses a 4-stage pipeline for these CISC instructions, where each instruction executes in a separate stage. The plan was to hide the execution of the other instructions by overlapping their execution with the `MatrixMultiply` instruction. Toward that end, the `Read_Weights` instruction follows the decoupled-access/execute philosophy [Smi82], in that it can complete after sending its address but before the weight is fetched from Weight Memory. The matrix unit will stall if the input activation or weight data is not ready.

We don't have clean pipeline overlap diagrams, because our CISC instructions can occupy a station for thousands of clock cycles, unlike the traditional RISC pipeline with one clock cycle per stage. Interesting cases occur when the activations for one network layer must complete before the matrix multiplications of the next layer can begin; we see a "delay slot," where the matrix unit waits for explicit synchronization before safely reading from the Unified Buffer.

As reading a large SRAM uses much more power than arithmetic, the matrix unit uses systolic execution to save energy by reducing reads and writes of the Unified Buffer [Kun80][Ram91][Ovt15b]. It relies on data from different directions arriving at cells in an array at regular intervals where they are combined. Figure 4 shows that data flows in from the left, and the weights are loaded from the top. A given 256-element multiply-accumulate operation moves through the matrix as a diagonal wavefront. The weights are preloaded, and take effect with the advancing wave alongside the first data of a new block. Control and data are pipelined to give the illusion that the 256 inputs are read at once, and that they instantly update one location of each of 256 accumulators. From a correctness perspective, software is unaware of the systolic nature of the matrix unit, but for performance, it does worry about the latency of the unit.

The TPU software stack had to be compatible with those developed for CPUs and GPUs so that applications could be ported quickly to the TPU. The portion of the application run on the TPU is typically written in TensorFlow and is compiled into an API that can run on GPUs or TPUs [Lar16]. Like GPUs, the TPU stack is split into a User Space Driver and a Kernel Driver. The Kernel Driver is lightweight and handles only memory management and interrupts. It is designed for long-term stability. The User Space driver changes frequently. It sets up and controls TPU execution, reformats data into TPU order, translates API calls into TPU instructions, and turns them into an application binary. The User Space driver compiles a model the first time it is evaluated, caching the program image and writing the weight image into the TPU's weight memory; the second and following evaluations run at full speed. The TPU runs most models completely from inputs to outputs, maximizing the ratio of TPU compute time to I/O time. Computation is often done one layer at a time, with overlapped execution allowing the matrix multiply unit to hide most non-critical-path operations.

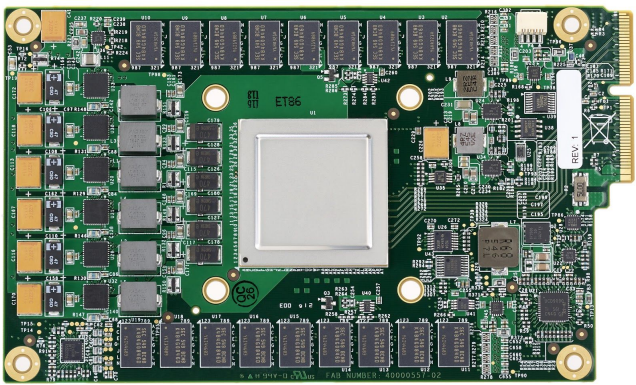

**Figure 3.** TPU Printed Circuit Board. It can be inserted in the slot for an SATA disk in a server, but the card uses PCIe Gen3 x16.

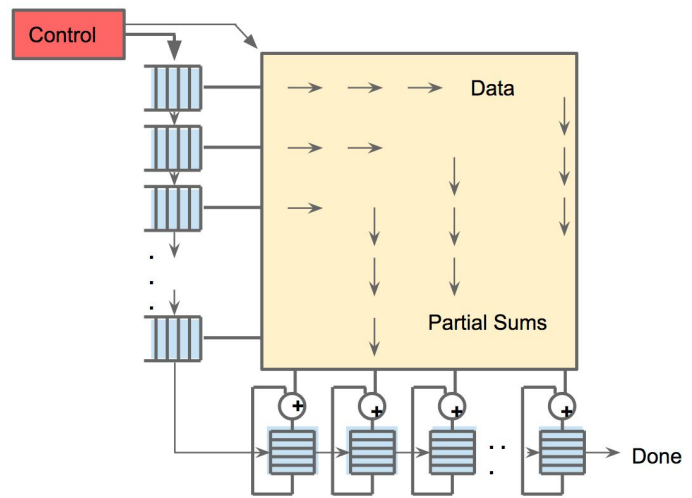

**Figure 4.** Systolic data flow of the Matrix Multiply Unit. Software has the illusion that each 256B input is read at once, and they instantly update one location of each of 256 accumulator RAMs.

| Model | Die | | | | | | | | | Benchmarked Servers | | | |
|---|---|---|---|---|---|---|---|---|---|---|---|---|---|
| | mm² | nm | MHz | TDP | Measured | | TOPS/s | | GB/s | On-Chip Memory | Dies | DRAM Size | TDP | Measured |
| | | | | | Idle | Busy | 8b | FP | | | | | | Idle | Busy |
| Haswell E5-2699 v3 | 662 | 22 | 2300 | 145W | 41W | 145W | 2.6 | 1.3 | 51 | 51 MiB | 2 | 256 GiB | 504W | 159W | 455W |
| NVIDIA K80 (2 dies/card) | 561 | 28 | 560 | 150W | 25W | 98W | -- | 2.8 | 160 | 8 MiB | 8 | 256 GiB (host) + 12 GiB x 8 | 1838W | 357W | 991W |
| TPU | NA* | 28 | 700 | 75W | 28W | 40W | 92 | -- | 34 | 28 MiB | 4 | 256 GiB (host) + 8 GiB x 4 | 861W | 290W | 384W |

**Table 2.** Benchmarked servers use Haswell CPUs, K80 GPUs, and TPUs. Haswell has 18 cores, and the K80 has 13 SMX processors. Figure 10 has measured power. The low-power TPU allows for better rack-level density than the high-power GPU. The 8 GiB DRAM per TPU is Weight Memory. GPU Boost mode is not used (Sec. 8). SECDEC and no Boost mode reduce K80 bandwidth from 240 to 160. No Boost mode and single die vs. dual die performance reduces K80 peak TOPS from 8.7 to 2.8. (*The TPU die is ≤ half the Haswell die size.)

### 3. CPU, GPU, and TPU Platforms

The six production applications in Table 1 are our workload for this paper. As mentioned above, these six are representative of 95% of TPU use in our datacenters. Ironically, deploying and measuring popular small DNNs like AlexNet or VGG is difficult on production machines. However, one of our CNNs derives from Inception V2, which is widely used.

The benchmark platforms are server-class computers that were available in 2015 when the TPUs were deployed. This restriction meant that they must include at least SECDED protection of internal SRAM as well as external DRAM memory like the TPU, which excludes some choices such as the Nvidia Maxwell GPU. For our company to purchase and deploy them, they also had to be sensibly configured machines, and not awkward artifacts assembled solely to win benchmarks.

Table 2 lists our choices. The traditional CPU server is represented by an 18-core, dual-socket Haswell processor from Intel. This platform is also the host server for GPUs or TPUs. Haswell was fabbed in an Intel 22nm process. Both the CPU and GPU are very large dies: about 600 mm$^2$!

The 2.3 GHz CPU clock rate doesn't include Turbo mode because it seldom occurs in our datacenters for NN apps. Haswell has different clock rates depending on whether programs use AVX instructions, which our NN apps often use. The higher clock rate of Turbo mode (for programs that avoid AVX) occurs when they don't use all their cores. Thus, another reason Turbo mode is rare in our datacenters is that our apps typically do use all the cores, plus they can run other datacenter jobs to fill any idle cores.

The GPU accelerator is the Nvidia K80. Each K80 card contains two dies and offers SECDED on internal memory and DRAM. Nvidia states that the "K80 Accelerator dramatically lowers datacenter cost by delivering application performance with fewer, more powerful servers"[Nvi16]. NN researchers frequently used K80s in 2015, and they were chosen for new cloud-based GPU offerings as recently as September 2016 [Bar16].

Up to eight K80 dies can be installed in four cards on this server, which is the configuration we benchmark. It offers a Boost mode to increase clock rate as high as 875 MHz. Turbo mode in Haswell is controlled by hardware and so can operate in short bursts before the chip temperature rises significantly. However, Boost mode is under the control of a software driver [Nvi15], and thus lasts at least hundreds of milliseconds. Hence, power and cooling would have to be provisioned for the K80 as if it were essentially always running in Boost mode, for otherwise the chips could get too hot. For this platform, enabling Boost mode would force us to reduce the number of K80 cards, which would hurt total cost of ownership. Thus, Boost mode is disabled. This restriction reduces peak bandwidth and TOPS (see Table 2 caption); Sec. 8 examines if it were enabled.

As the number of dies per benchmarked server varies between 2 to 8, we usually show results below normalized per die (Figures 5-8, Figures 10-11, and Tables 3, 4, and 6), but we occasionally show whole systems (Figure 9). We hope this distinction is clear.

### 4. Performance: Rooflines, Response-Time, and Throughput

To illustrate the performance of the six apps on the three processors, we adapt the Roofline Performance model from high-performance computing (HPC) [Wil09]. This simple visual model is not perfect, yet it offers insights on the causes of performance bottlenecks. The assumption behind the model is that applications don't fit in on-chip caches, so they are either computation-limited or memory bandwidth-limited. For HPC, the Y-axis is performance in floating-point operations per second, thus the peak computation rate forms the "flat" part of the roofline. The X-axis is operational intensity, measured as floating-point operations per DRAM byte accessed. Memory bandwidth is bytes per second, which turns into the "slanted" part of the roofline since (FLOPS/sec)/ (FLOPS/Byte) = Bytes/sec. Without sufficient operational intensity, a program is memory bandwidth-bound and lives under the slanted part of the roofline.

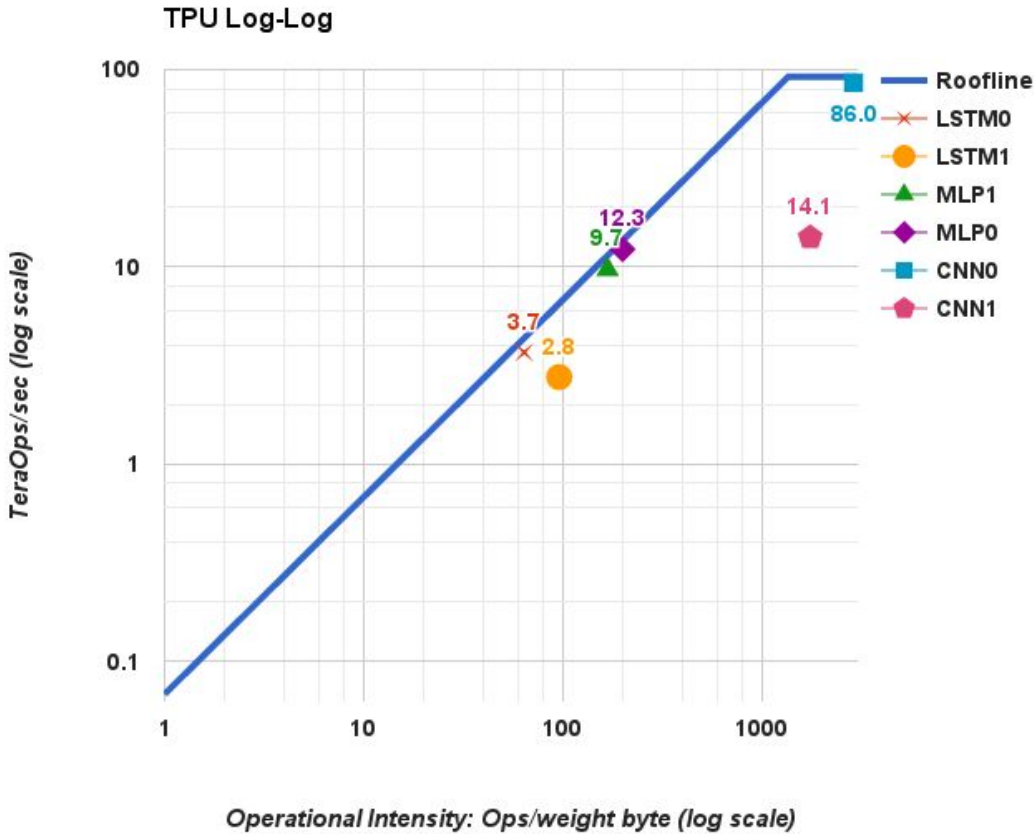

**Figure 5.** TPU (die) roofline. Its ridge point is far to the right at 1350 operations per byte of weight memory fetched.

The gap between the actual operations per second of an application and the ceiling directly above it shows the potential benefit of further performance tuning while leaving operational intensity untouched; of course, optimizations that increase operational intensity (such as cache blocking) may yield even greater benefit.

To use the Roofline model for the TPU, when NN applications are quantized, we first replace floating-point operations with integer operations. As weights do not normally fit in on-chip memory for NN applications, the second change is to redefine operational intensity to be integer operations per byte of *weights* read (see the tenth column of Table 1).

Figure 5 shows the Roofline model for a single TPU die on log-log scales. The TPU has a long "slanted" part of its roofline, where operational intensity means that performance is limited by memory bandwidth rather than by peak compute. Five of the six applications are happily bumping their heads against the ceiling: the MLPs and LSTMs are memory bound, and CNNs are computation bound. CNN1, despite a very high operational intensity, is running at only 14.1 TOPS while CNN0 runs at 86 TOPS.

Table 3 explains what happened with CNN1, based on the performance counters that give us partial visibility into TPU operation. The TPU spends less than half of its cycles performing matrix operations for CNN1 (column 7, row 1). On each of those active cycles, only about half of the 65,536 MACs hold useful weights because some layers in CNN1 have shallow feature depths. About 35% of cycles are spent waiting for weights to load from memory into the matrix unit, which occurs during the 4 fully connected layers that run at an operational intensity of just 32 (see the last Fallacy in Section 8). This leaves roughly 19% of cycles not explained by the matrix-related counters. Because of overlapped execution on the TPU, we do not have exact accounting for those cycles, but we can see that 23% of cycles have stalls for RAW dependences in the pipeline, and 1% are spent stalled for input over the PCIe bus.

| Application | MLP0 | MLP1 | LSTM0 | LSTM1 | CNN0 | CNN1 | Mean | Row |
|---|---|---|---|---|---|---|---|---|
| Array active cycles | 12.7% | 10.6% | 8.2% | 10.5% | 78.2% | 46.2% | 28% | 1 |
|    Useful MACs in 64K matrix (% peak) | 12.5% | 9.4% | 8.2% | 6.3% | 78.2% | 22.5% | 23% | 2 |
|    Unused MACs | 0.3% | 1.2% | 0.0% | 4.2% | 0.0% | 23.7% | 5% | 3 |
| Weight stall cycles | 53.9% | 44.2% | 58.1% | 62.1% | 0.0% | 28.1% | 43% | 4 |
| Weight shift cycles | 15.9% | 13.4% | 15.8% | 17.1% | 0.0% | 7.0% | 12% | 5 |
| Non-matrix cycles | 17.5% | 31.9% | 17.9% | 10.3% | 21.8% | 18.7% | 20% | 6 |
| RAW stalls | 3.3% | 8.4% | 14.6% | 10.6% | 3.5% | 22.8% | 11% | 7 |
| Input data stalls | 6.1% | 8.8% | 5.1% | 2.4% | 3.4% | 0.6% | 4% | 8 |
| TeraOps/sec (92 Peak) | 12.3 | 9.7 | 3.7 | 2.8 | 86.0 | 14.1 | 21.4 | 9 |

**Table 3.** Factors limiting TPU performance of the NN workload based on hardware performance counters. Rows 1, 4, 5, and 6 total 100% and are based on measurements of activity of the matrix unit. Rows 2 and 3 further break down the fraction of 64K weights in the matrix unit that hold useful weights on active cycles. Our counters cannot exactly explain the time when the matrix unit is idle in row 6; rows 7 and 8 show counters for two possible reasons, including RAW pipeline hazards and PCIe input stalls. Row 9 (TOPS) is based on measurements of production code while the other rows are based on performance-counter measurements, so they are not perfectly consistent. Host server overhead is excluded here. The MLPs and LSTMs are memory-bandwidth limited but CNNs are not. CNN1 results are explained in the text.

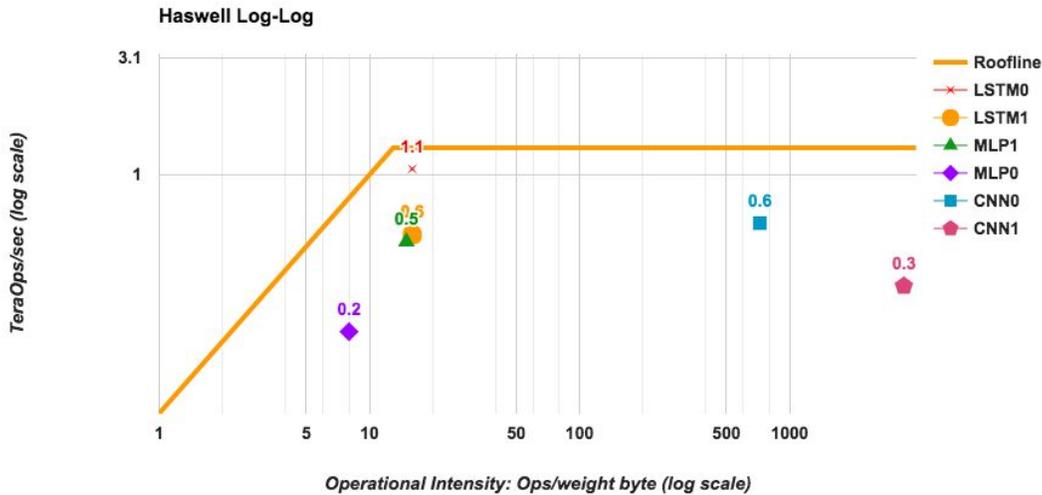

**Figure 6.** Intel Haswell CPU (die) roofline with its ridge point at 13 operations/byte, which is much further left than in Figure. 5. LSTM0 and MLP1 are faster on Haswell than on the K80, but it is vice versa for the other DNNs.

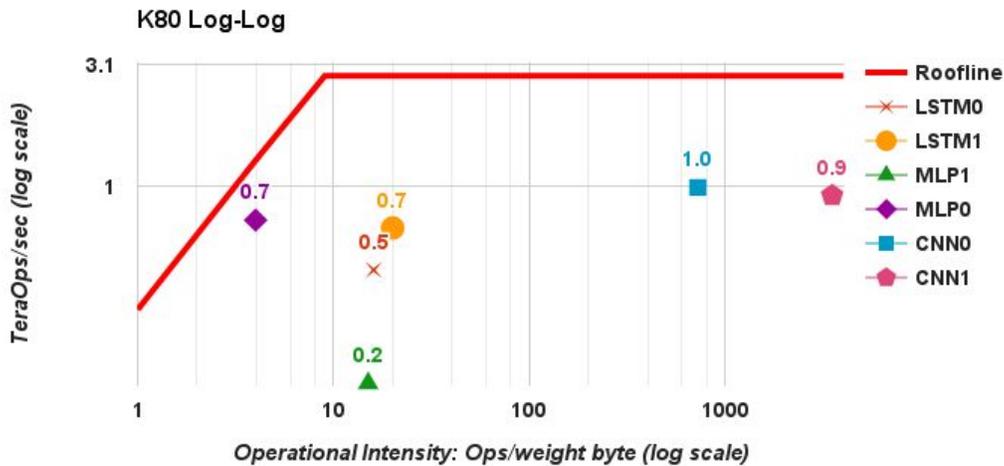

**Figure 7.** NVIDIA K80 GPU die Roofline. The much higher memory bandwidth moves the ridge point to 9 operations per weight byte, which is even further left than in Figure 6. The DNNs are much lower than their Roofline because of response time limits (see Table 4).

Figures 6 and 7 show rooflines for a single Haswell die and for a single K80 die. The six NN applications are generally further below their ceilings than was the TPU in Figure 5. Response time is the reason. Many of these NN applications are parts of end-user-facing services. Researchers have demonstrated that small increases in response time cause customers to use a service less [Sch09]. Hence, while training may not have hard response time deadlines, inference usually does. That is, inference prefers latency over throughput [Pat04].

Table 4 illustrates the impact of the 99th-percentile response time limit of 7 ms for MLP0 on Haswell and the K80 [Dea13], which was required by the application developer. (The inferences per second and 7 ms latency include the server host time as well as the accelerator time.) They operate at 42% and 37%, respectively, of the highest throughput achievable for MLP0 if the response time limit was relaxed. Thus, while CPUs and GPUs have potentially much higher throughput, it's wasted if they don't meet the response time limit. These bounds affect the TPU as well, but at 80% in Table 4 it is operating much closer to its highest MLP0 throughput. As compared to CPUs and GPUs, the single-threaded TPU has none of the sophisticated microarchitectural features that consume transistors and energy to improve the average case but not the 99th-percentile case: no caches, branch prediction, out-of-order execution, multiprocessing, speculative prefetching, address coalescing, multithreading, context switching, and so forth. Minimalism is a virtue of domain-specific processors.

Table 3 shows TPU performance, but it doesn't account for host server time, which can be divided into running the host share of the application and talking to the TPU. Table 5 lists the second part, but the first part is hard. Queueing theory shows that long input queues raise throughput—by ensuring that the computer is never idle—but stretch response time. Thus, most applications keep their input queues empty. Alas, we can't measure when the TPU is idle since it is waiting for the CPU to do its portion of the application or because the CPU is also idle due to an empty input queue.

Table 6 gives the bottom line of relative inference performance per die including the host server overhead for the two accelerators versus the CPU. The next-to-last column shows the geometric mean of the relative performance for the six NN applications, which suggests the K80 die is 1.1X the speed of a Haswell die, that the TPU die is 14.5 times as fast, and thus the TPU die is 13.2 times as fast as the GPU die. Figure 8 shows their relative speeds visually.

Recall that architects use the geometric mean when they don't know the actual mix of programs that will be run [Hen18]. For this study, however, we *do* know the mix (Table 1). The *weighted* mean in the last column of Table 6 using the actual mix increases the GPU to 1.9X and the TPU to 29.2X, so the TPU die is now 15.3 times as fast as the GPU die.

| Type | Batch | 99th% Response | Inf/s (IPS) | % Max IPS |
|------|-------|----------------|-------------|-----------|
| CPU  | 16    | 7.2 ms         | 5,482       | 42%       |
| CPU  | 64    | 21.3 ms        | 13,194      | 100%      |
| GPU  | 16    | 6.7 ms         | 13,461      | 37%       |
| GPU  | 64    | 8.3 ms         | 36,465      | 100%      |
| TPU  | 200   | 7.0 ms         | 225,000     | 80%       |
| TPU  | 250   | 10.0 ms        | 280,000     | 100%      |

**Table 4.** 99-th% response time and per die throughput (IPS) for MLP0 as batch size varies for MLP0. The longest allowable latency is 7 ms. For the GPU and TPU, the maximum MLP0 throughput is limited by the host server overhead. Larger batch sizes increase throughput, but as the text explains, their longer response times exceed the limit, so CPUs and GPUs must use less-efficient, smaller batch sizes (16 vs. 200).

| MLP0 | MLP1 | LSTM0 | LSTM1 | CNN0 | CNN1 |
|------|------|-------|-------|------|------|
| 21%  | 76%  | 11%   | 20%   | 51%  | 14%  |

**Table 5.** Time for host CPU to interact with the TPU expressed as percent of TPU execution time (from TPU performance counters). This fraction is the time the CPU and TPU are communicating over the PCIe bus, *not* including the time the CPU is doing a portion of the application but not interacting with the TPU. As the text explains, it's hard for the TPU to measure if the CPU is idle or working on the app.

| Type  | MLP0 | MLP1 | LSTM0 | LSTM1 | CNN0 | CNN1 | GM   | WM   |
|-------|------|------|-------|-------|------|------|------|------|
| GPU   | 2.5  | 0.3  | 0.4   | 1.2   | 1.6  | 2.7  | 1.1  | 1.9  |
| TPU   | 41.0 | 18.5 | 3.5   | 1.2   | 40.3 | 71.0 | 14.5 | 29.2 |
| Ratio | 16.7 | 60.0 | 8.0   | 1.0   | 25.4 | 26.3 | 13.2 | 15.3 |

**Table 6.** K80 GPU die and TPU die performance relative to CPU for the NN workload. GM and WM are geometric and weighted mean (using the actual mix of the six apps in Table 1). Relative performance for the GPU and TPU includes host server overhead. MLPs and CNNs perform well on the TPU. Table 4 explains that the TPU can have larger batch sizes and still meet the time limits, increasing operations per byte (Table 1) or, equivalently, reducing memory accesses per operation. Also, CNNs by their nature have greater weight reuse and thus higher operations per byte. Thus, the lower memory bandwidth of the TPU doesn't significantly hurt CNN performance.

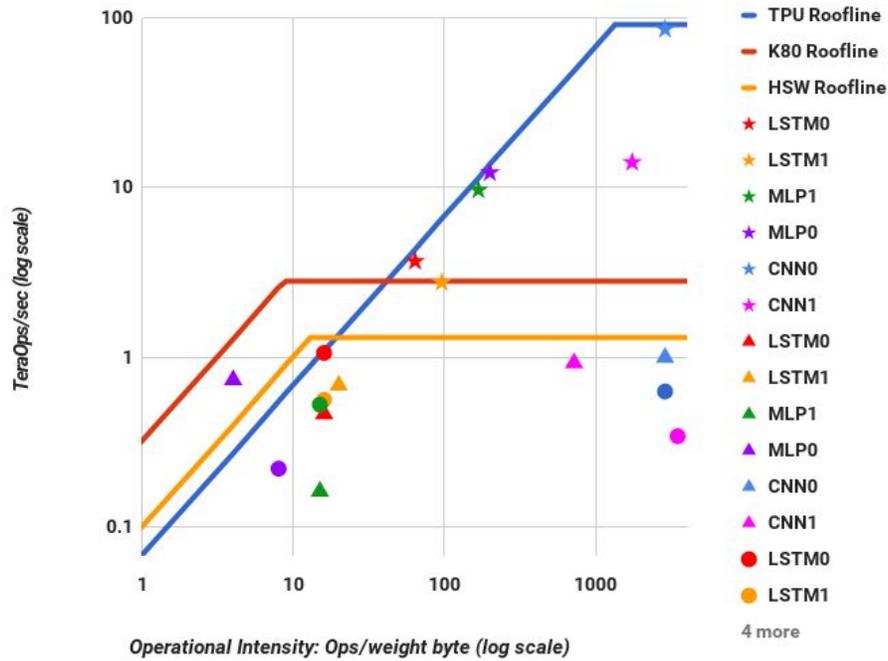

**Figure 8.** Figures 5-7 combined into a single log-log graph. Stars are for the TPU, triangles are for the K80, and circles are for Haswell. All TPU stars are at or above the other 2 rooflines.

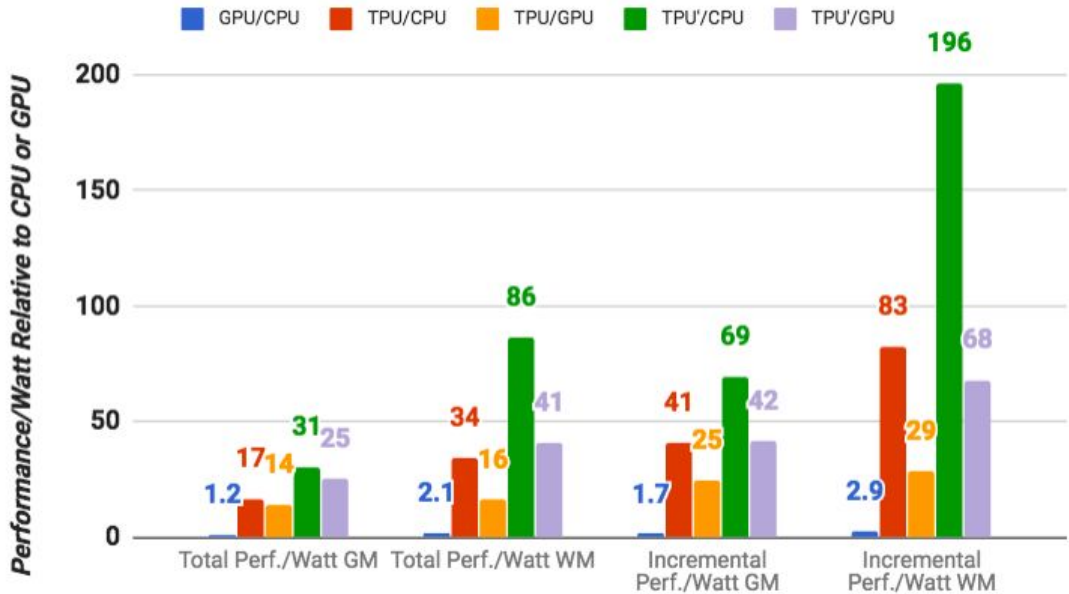

**Figure 9.** Relative performance/Watt (TDP) of GPU server (blue bar) and TPU server (red bar) to CPU server, and TPU server to GPU server (orange bar). TPU' is an improved TPU (Sec. 7). The green bar shows its ratio to the CPU server and the lavender bar shows its relation to the GPU server. Total includes host server power, but incremental doesn't. GM and WM are the geometric and weighted means.

## 5. Cost-Performance, TCO, and Performance/Watt

When buying computers by the thousands, cost-performance trumps performance. The best cost metric in a datacenter is total cost of ownership (TCO). The actual price we pay for thousands of chips depends on negotiations between the companies involved. For business reasons, we can't publish such price information or data that might let them be deduced. However, power is correlated with TCO, and we *can* publish Watts per server, so we use performance/Watt as our proxy for performance/TCO in this paper. In this section, we compare whole servers rather than single dies, which Table 2 lists in the "Benchmarked Server" columns.

Figure 9 shows the geometric and weighted mean performance/Watt for the K80 GPU and TPU relative to the Haswell CPU. We present two different calculations of performance/Watt. The first ("total") includes the power consumed by the host CPU server when calculating performance/Watt for the GPU and TPU. The second ("incremental") subtracts the host CPU server power from the GPU and TPU beforehand.

For *total*-performance/Watt, the K80 server is 1.2 - 2.1X Haswell. For *incremental*-performance/Watt, when Haswell server power is omitted, the K80 server is 1.7- 2.9X.

The TPU server has 17 to 34 times better total-performance/Watt than Haswell, which makes the TPU server 14 to 16 times the performance/Watt of the K80 server. The relative incremental-performance/Watt—which was our company's justification for a custom ASIC—is 41 to 83 for the TPU, which lifts the TPU to 25 to 29 times the performance/Watt of the GPU.

## 6. Energy Proportionality

Thermal Design Power (TDP) affects the cost of provisioning power, as you must supply sufficient power and cooling when hardware is at full power. However, the cost of electricity is based on the *average* consumed as the workload varies during the day. [Bar07] found that servers are 100% busy less than 10% of the time and advocated *energy proportionality*: servers should consume power proportional to the amount of work performed. The estimate of power consumed in the prior section is based on the fraction of the TDP that has been seen in our datacenters.

We measured performance and power as the offered workload utilization varies from 0% to 100%, collected in buckets of 10% delta of workload [Lan09]. Figure 10 shows server power divided by the number of dies per server for the three chips by varying CNN0's workload. We plot incremental (K80 and TPU) as well as total power (K80+Haswell/4 and TPU+Haswell/2) for the GPU and TPU. Note that all were given the same batch sizes.

We see that the TPU has the lowest power—118W per die total (TPU+Haswell/2) and 40W per die incremental (TPU in Fig. 10)— but it has poor energy proportionality: at 10% load, the TPU uses 88% of the power it uses at 100%. (The short design schedule prevented inclusion of many energy-saving features.) Not surprisingly, Haswell is the best at energy proportionality of the group: it uses 56% of the power at 10% load as it does at 100%. The K80 is closer to the CPU than the TPU, using 66% of the full load power at 10% workload. LSTM1, which is not computation bound, performs similarly: at 10% load the CPU uses 47% of full power, the GPU uses 78%, and the TPU uses 94%.

What happens to the server power usage when running CNN0 if it becomes a host to accelerators? When the GPU and TPU are at 100% load, the CPU server uses 52% of full power for the GPU and 69% for the TPU. (The CPU does more work for the TPU because it is running so much faster than the GPU.) Consequently, the Haswell server plus four TPUs use <20% additional power but run CNN0 80 times faster than the Haswell server alone (4 TPUs vs 2 CPUs).

## 7. Evaluation of Alternative TPU Designs

Like an FPU, the TPU coprocessor has a relatively easy microarchitecture to evaluate, so we created a performance model for our six applications. Table 7 shows the differences between the model results and the hardware performance counters, which average below 10%. We then modeled performance as we varied the memory bandwidth, the clock rate and number of accumulators, and the matrix multiply unit size.

Figure 11 shows the mean performance sensitivity of TPU die as we scale these parameters over the range for 0.25x to 4x. It plots weighted means, but the geometric means look similar. In addition to evaluating the impact of only raising clock rates (clock in Figure 11), we also plot a design (clock+) where the clock rate is increased *and* the number of accumulators is correspondingly scaled so the compiler can keep more memory references in flight. Likewise, we plot matrix unit expansion if we increase the number of accumulators with the *square* of the rise in one dimension (matrix+), since the number of multipliers in the matrix grows in both dimensions, as well as just increasing the matrix unit alone (matrix).

| MLP0 | MLP1 | LSTM0 | LSTM1 | CNN0 | CNN1 |
|------|------|-------|-------|------|------|
| 6.8% | 10.9% | 7.7% | 5.4% | 8.2% | 11.2% |

**Table 7.** Difference in clock cycles between the TPU hardware performance counters and the TPU performance model. The average is 8%.

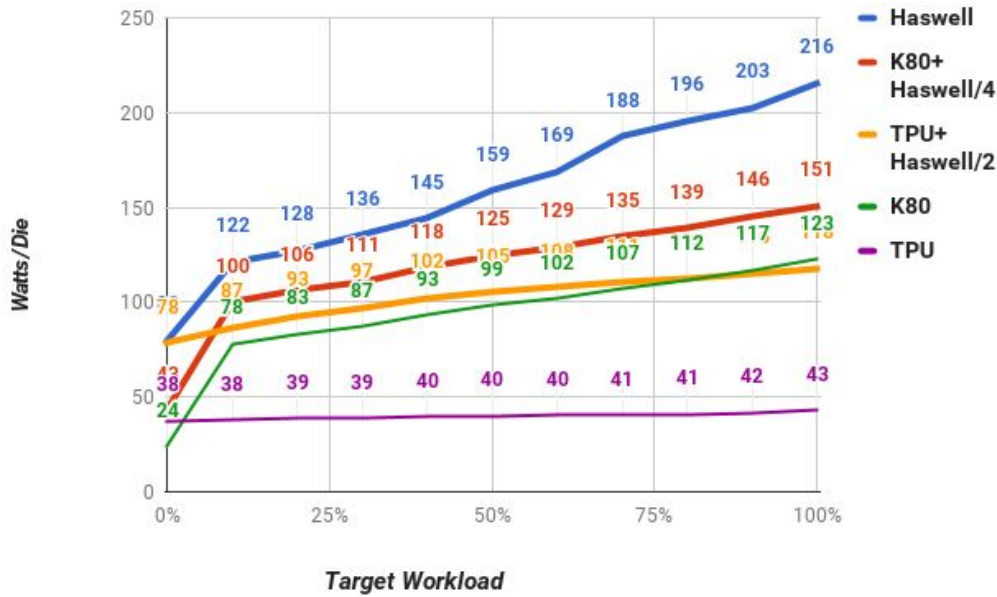

**Figure 10.** Watts/die for CNN0 as target platform utilization varies from 0% to 100%. The Total GPU and TPU power are the red and orange lines and their Incremental power are the green and purple lines. (The blue line is power for the Haswell CPU, which by definition is Total power.) A server has 2 CPUs and 8 GPUs or 4 TPUs, so we normalize power by dividing by 2, 8, and 4, respectively.

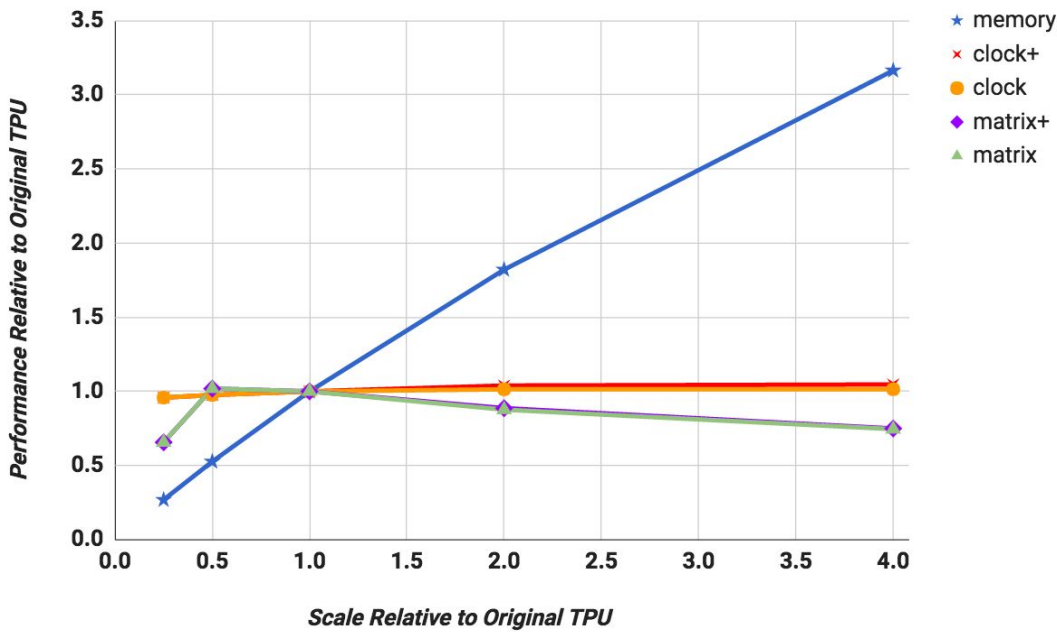

**Figure 11.** Weighted mean TPU performance as metrics scale from 0.25x to 4x: memory bandwidth, clock rate + accumulators, clock rate, matrix unit dimension + accumulators, and matrix unit dimension. The weighted mean makes it hard see to contributions of individual DNNs, but MLPs and LSTMs improve 3X with 4X memory bandwidth, but get nothing from a higher clock. For CNNs it's vice versa; 2X for 4X clock, but get little benefit from faster memory. A bigger matrix multiply unit doesn't help any DNN.

First, increasing memory bandwidth (**memory**) has the biggest impact: performance improves 3X on average when memory increases 4X. Second, clock rate has little benefit on average with or without more accumulators. The reason is the MLPs and LSTMs are memory bound but only the CNNs are compute bound. While hard to see in Figure 11, since it shows only the weighted mean of all six DNNs, increasing the clock rate by 4X has almost no impact on MLPs and LSTMs but improves performance of CNNs by about 2X. Third, the average performance in Figure 11 slightly *degrades* when the matrix unit expands from 256x256 to 512x512 for all apps, whether or not they get more accumulators. The issue is analogous to internal fragmentation of large pages, only worse since it's in two dimensions. Consider the 600x600 matrix used in LSTM1. With a 256x256 matrix unit, it takes 9 steps to tile 600x600, for a total of 18 us of time. The larger 512x512 unit requires only four steps, but each step takes four times longer, for 32 us of time. Our CISC instructions are long, so decode is insignificant and does not hide the overhead of loading from the DRAM.

Table 8 shows the utilization of the 24 MiB Unified Buffer, which was initially sized to allow MLPs to run at batch sizes up to 2048. We recently improved the storage allocator for the Unified Buffer, which reduces the memory needed for the largest of the six applications to 14 MiB. For the first 18 months of deployment, the TPU used its full capacity while the new allocator was being developed. Now the extra capacity adds margin for adopting bigger models.

We next used the performance model to evaluate a hypothetical TPU die (*TPU'*) that might have been designed in the same process technology if we had more than 15 months. More aggressive logic synthesis and block design might have increased the clock rate by 50%. Designing an interface circuit for GDDR5 memory, as in the K80, would improve Weight Memory bandwidth by more than a factor of *five*, shifting its roofline ridge point from 1350 to 250. As Figure 11 shows, increasing clock rate to 1050 MHz but not helping memory makes almost no change. If we left the clock at 700 MHz but used GDDR5 for Weight Memory, the geometric mean increase jumps to an impressive 2.6 and the weighted mean to 3.9. Doing both raises the geometric mean (2.9) but not the weighted mean, so TPU' just has faster memory.

Figure 11 does *not* include host server time. We used Table 5 to calculate time for the host server interaction overhead for the TPU. Adding that same extra time drops TPU' means from 2.6 to 1.9 and 3.9 to 3.2. This change is both optimistic, since it doesn't include CPU time to run its share of the app, and pessimistic, as we likely would aggressively tune the host code given a 3X faster TPU'.

Replacing just the DDR3 Weight Memory with the equivalent GDDR5 memory requires doubling the number of memory channels to four. This improvement would expand die size by about 10%. However, higher memory bandwidth reduces pressure on the Unified Buffer, so reducing the Unified Buffer to 14 MiB could gain back 10% in area. GDDR5 would also increase the TPU system power budget from 861 Watts to about 900 Watts, as there are 4 TPUs per server.

Figure 9 above shows the relative total-performance/Watt/die of TPU' leaps to 31X - 86X over Haswell and 25X - 41X over the K80. The incremental metric soars to an amazing 69X - 196X over Haswell and 42X - 68X over the K80.

| MLP0 | MLP1 | LSTM0 | LSTM1 | CNN0 | CNN1 |
|---|---|---|---|---|---|
| 11.0 | 2.3 | 4.8 | 4.5 | 1.5 | 13.9 |

**Table 8.** Maximum MiB of the 24 MiB Unified Buffer used per NN app. A 14 MiB Unified Buffer is sufficient for these apps.

## 8. Discussion

This section follows the fallacy and pitfall with rebuttal style of [Hen18].

- *Fallacy: NN inference applications in datacenters value throughput as much as response time.*

We were surprised that our developers had strong response-time demands, as some suggested in 2014 that batch sizes would be large enough for the TPU to reach peak performance or that latency requirements wouldn't be as tight. One driving application was off-line image processing, and the intuition was that if interactive services also wanted TPUs, most of them would just accumulate larger batches. Even the developers of one application in 2014 that cared about response time (LSTM1) said the limit was 10 ms in 2014, but shrank it to 7 ms when they actually ported it to the TPU. The unexpected desire for TPUs by many such services combined with the impact on and preference for low response time changed the equation, with application writers often opting for reduced latency over waiting for bigger batches to accumulate. Fortunately, the TPU has a simple and repeatable execution model to help meet the response-time targets of interactive services and such high peak throughput that even small batch sizes result in higher performance than contemporary CPUs and GPUs.

- *Fallacy: The K80 GPU architecture is a good match to NN inference.*

GPUs have traditionally been seen as high-throughput architectures that rely on high-bandwidth DRAM and thousands of threads to achieve their goals. This perspective helps explain why the K80 is only a little faster at inference than Haswell and much slower than the TPU. Successors to the K80 will surely include optimizations to improve peak inference performance, but given their throughput-oriented architectural approach, it may be more challenging for GPUs to meet the strict latency limits. And as Section 7 shows, there is plenty of headroom to improve the TPU, so it's not an easy target.

- *Pitfall: Architects have neglected important NN tasks.*

We are pleased by the attention that the architecture community is paying to NN: 15% of the papers at ISCA 2016 were on hardware accelerators for NN [Alb16] [Che16a][Chi16][Han16][Kim16][LiK16][Liu16][Rea16] [Sha16]! Alas, all nine papers looked at CNNs, and only two mentioned other NNs. CNNs are more complex than MLPs and prominent in NN competitions [Rus15], which might explain their allure, but they are only about 5% of our datacenter NN workload. While CNNs may be common in edge devices, the volume of convolutional models hasn't yet caught up with MLPs and LSTMs in the datacenter. We hope that architects try to accelerate MLPs and LSTMs with at least as much gusto.

- *Pitfall: For NN hardware, Inferences Per Second (IPS) is an inaccurate summary performance metric.*

Our results show that IPS is a poor overall performance summary for NN hardware, as it's simply the inverse of the complexity of the typical inference in the application (e.g., the number, size, and type of NN layers). For example, the TPU runs the 4-layer MLP1 at 360,000 IPS but the 89-layer CNN1 at only 4,700 IPS, so TPU IPS vary by 75X! Thus, using IPS as the single-speed summary is *even more misleading* for NN accelerators than MIPS or FLOPS are for regular processors [Hen18], so IPS should be even more disparaged. To compare NN machines better, we need a benchmark suite written at a high-level to port it to the wide variety of NN architectures. Fathom is a promising new attempt at such a benchmark suite [Ado16].

- *Fallacy: The K80 GPU results would be much better if Boost mode were enabled.*

Setting aside the negative impact of K80 Boost mode on TCO (Section 3), we measured it on LSTM1. Boost mode increased the clock rate by a factor of up to 1.6—from 560 to 875 MHz—which increased performance by 1.4X, but it also raised power by 1.3X. The net gain in performance/Watt is 1.1X, and thus for LSTM1, boost mode would have a minor impact on our energy-speed analysis.

- *Fallacy: CPU and GPU results would be comparable to the TPU if we used them more efficiently or compared to newer versions.*

We originally had 8-bit results for just one DNN on the CPU, due to the significant work to use AVX2 integer support efficiently. The benefit was ~3.5X. It was less confusing (and less space) to present all CPU results in floating point, rather than having one exception, with its own roofline. If all DNNs had similar speedup, performance/Watt ratio would drop from 41-83X to 12-24X. The new 16-nm, 1.5GHz, 250W P40 datacenter GPU can perform 47 Tera 8-bit ops/sec, but was unavailable in early 2015, so isn't contemporary with our three platforms. We also can't know the fraction of P40 peak delivered within our rigid time bounds. If we compared newer chips, Section 7 shows that we could triple performance of the 28-nm, 0.7GHz, 40W TPU just by using the K80's GDDR5 memory (at a cost of an additional 10W).

- *Pitfall: Performance counters added as an afterthought for NN hardware.*

The TPU has 106 performance counters, and if anything we would like a few more (see Table 3). The raison d'etre for NN accelerators is performance, and it is way too early in their evolution to have good intuition about what is going on.

- *Fallacy: After two years of software tuning, the only path left to increase TPU performance is hardware upgrades.*

The performance of CNN1 on the TPU could improve if developers and compiler writers did more work to match CNN1 to the TPU hardware. For example, developers could reorganize the applications to aggregate multiple short batches out of the convolution layers into a single, deeper batch (from 32 to 128) for the four fully connected layers. Such a single layer would improve utilization of the matrix unit (Table 3). As CNN1 currently runs more than 70 times faster on the TPU than the CPU, the CNN1 developers are already very happy, so it's not clear whether or when such optimizations would be performed.

9. **Related Work**

Two survey articles document that custom NN ASICs go back at least 25 years [Ien96][Asa02]. For example, CNAPS chips contained a 64 SIMD array of 16-bit by 8-bit multipliers, and several CNAPS chips could be connected together with a sequencer [Ham90]. The Synapse-1 system was based on a custom systolic multiply-accumulate chip called the MA-16, which performed sixteen 16-bit multiplies at a time [Ram91]. The system concatenated several MA-16 chips together and had custom hardware to do activation functions.

Twenty-five SPERT-II workstations, accelerated by the T0 custom ASIC, were deployed starting in 1995 to do both NN training and inference for speech recognition [Asa98]. The 40-Mhz T0 added vector instructions to the MIPS instruction set architecture. The eight-lane vector unit could produce up to sixteen 32-bit arithmetic results per clock cycle based on 8-bit and 16-bit inputs, making it 25 times faster at inference and 20 times faster at training than a SPARC-20 workstation. They found that 16 bits were insufficient for training, so they used two 16-bit words instead, which doubled training time. To overcome that drawback, they introduced "bunches" (batches) of 32 to 1000 data sets to reduce time spent updating weights, which made it faster than training with one word but no batches.

The more recent DianNao family of NN architectures minimizes memory accesses both on the chip and to external DRAM by having efficient architectural support for the memory access patterns that appear in NN applications [Keu16] [Che16a]. All use 16-bit integer operations and all designs dove down to layout, but no chips were fabricated. The original DianNao uses an array of 64 16-bit integer multiply-accumulate units with 44 KB of on-chip memory and is estimated to be 3 mm$^2$ (65 nm), to run at 1 GHz, and to consume 0.5W [Che14a]. Most of this energy went to DRAM accesses for weights, so one successor DaDianNao ("big computer") includes eDRAM to keep 36 MiB of weights on chip [Che14b]. The goal was to have enough memory in a multichip system to avoid external DRAM accesses. The follow-on PuDianNao ("general computer") is aimed at more traditional machine learning algorithms beyond DNNs, such as support vector machines [Liu15]. Another offshoot is ShiDianNao ("vision computer") aimed at CNNs, which avoids DRAM accesses by connecting the accelerator directly to the sensor [Du15].

The Convolution Engine is also focused on CNNs for image processing [Qad13]. This design deploys 64 10-bit multiply-accumulator units and customizes a Tensilica processor estimated to run at 800 MHz in 45 nm. It is projected to be 8X to 15X more energy-area efficient than an SIMD processor, and within 2X to 3X of custom hardware designed just for a specific kernel.

The Fathom benchmark paper seemingly reports results contradictory to ours, with the GPU running inference much faster than the CPU [Ado16]. However, their CPU and GPU are not server-class, the CPU has only four cores, the applications do not use the CPU's AVX instructions, and there is no response-time cutoff (see Table 4) [Bro16].

Catapult is the most widely deployed example of using reconfigurability to support DNNs, which many have proposed [Far09][Cha10][Far11][Pee13][Cav15][Zha15]. They chose FPGAs over GPUs to reduce power as well as the risk that latency-sensitive applications wouldn't map well to GPUs. FPGAs can also be re-purposed, such as for search, compression, and network interface cards [Put15]. The TPU project actually began with FPGAs, but we abandoned them when we saw that the FPGAs of that time were not competitive in performance compared to the GPUs of that time, and the TPU could be much lower power than GPUs while being as fast or faster, giving it potentially significant benefits over both of FPGAs and GPUs.

Although first published in 2014 [Put14], Catapult is a TPU contemporary since it deployed 28-nm Stratix V FPGAs into datacenters concurrently with the TPU in 2015. Catapult has a 200 MHz clock, 3,926 18-bit MACs, 5 MiB of on-chip memory, 11 GB/s memory bandwidth, and uses 25 Watts. The TPU has a 700 MHz clock, 65,536 8-bit MACs, 28 MiB, 34 GB/s, and typically uses 40 Watts. A revised version of Catapult uses newer FPGAs and was deployed at larger scale in 2016 [Cau 16].

Catapult V1 runs CNNs—using a systolic matrix multiplier—2.3X as fast as a 2.1 GHz, 16-core, dual-socket server [Ovt15a]. Using the next generation of FPGAs (14-nm Arria 10) of Catapult V2, performance might go up to 7X, and perhaps even 17X with more careful floorplanning [Ovt15b]. Although it's apples versus oranges, a current TPU die runs its CNNs 40X to 70X versus a somewhat faster server (Tables 2 and 6). Perhaps the biggest difference is that to get the best performance the user must write long programs in the low-level hardware-design-language Verilog [Met16][Put16] versus writing short programs using the high-level TensorFlow framework. That is, reprogrammability comes from software for the TPU rather than from firmware for the FPGA.

Recent research, which appeared after the TPU was deployed, accelerates DNNs by optimizing the cases when weights and data are very small or zero. Our tight schedule precluded such optimizations in the TPU, but we saw the same opportunity in our studies. The Efficient Inference Engine is based on a first pass that reduces the number of weights by about a factor of 10 [Han15] as a separate step by filtering out very small values and then uses Huffman encoding to shrink the data even further to improve inference performance [Han16]. Cnvlutin [Alb16] avoids multiplications when an activation input is zero—which it is 44% of the time, presumably in part due to ReLU nonlinear function that transforms negative values to zero—to improve performance by an average 1.4 times.

Eyeriss is a novel, low-power dataflow architecture that takes advantage of zeros by run-length encoding data to reduce the memory footprint and saves power by avoiding computations when an input is zero [Che16a]. Using Eyeriss terminology, a TPU convolutional layer maps C and M to the rows and columns of the matrix unit, taking HWN cycles to perform one pass. With high C/M, it takes RS passes to process the layer; for low C/M, a number of techniques reduce passes and improve utilization. (More can be found in the online references [Ros15a][Ros15b][Ros15c][Ros15f][Tho15][You15]).

Minerva is a co-design system that crosses algorithm, architecture, and circuit disciplines to reduce power by 8X in part by pruning activation data with small values and in part by quantizing the data [Rea16]. [Gup15] looks at 16-bit fixed-point arithmetic for training instead of for inference. Others leverage the lower precision of DNN calculations by utilizing analog circuits during the computation to improve energy and performance [LiK16] [Sha16]. By tailoring an instruction set to DNNs, Cambricon reduces code size [Liu16]. Recent work looked at processor-in-memory architectures for NNs [Chi16][Kim16].

Comparing the TPU to some of these architectures:
- [Che14a] DMAs data from DRAM to input and weight buffers. They are read by the 3-stage pipelined NFU that performs multiplies, adds, and non-linear-functions; the results go to the output buffer, and then to DRAM. The NFU has no storage and isn't systolic.

- [Gup15] appears to stream both matrix inputs while storing partial sums in the systolic array; the TPU stores the weight matrix tile while streaming the other input and the pre-activation partial sums. The TPU doesn't support stochastic rounding.
- [Zha15] is built out of computation units equivalent to a 4x2 version of the TPU matrix unit. In an ASIC, the wiring cost of the crossbars that connect input and output buffers to these compute engines would be significant. We are surprised that we didn't see architectural support for additional reductions to combine results from compute engines in [Zha15].

All three of [Gup15][Che14a][Zha15] store activations in DRAM during computation; the TPU's Unified Buffer is sized so that no DRAM spilling or reloading happens during normal operation.

## 10. Conclusion

Despite living on an I/O bus and having relatively little memory bandwidth that limits utilization of the TPU—four of the six NN applications are memory-bound—a small fraction of a big number can nonetheless be relatively large, as the Roofline performance model demonstrates. This result suggests a "Cornucopia Corollary" to Amdahl's Law: *low utilization of a huge, cheap resource can still deliver high, cost-effective performance*.

   The TPU leverages the order-of-magnitude reduction in energy and area of 8-bit integer systolic matrix multipliers over 32-bit floating-point datapaths of a K80 GPU to pack 25 times as many MACs (65,536 8-bit vs. 2,496 32-bit) and 3.5 times the on-chip memory (28 MiB vs. 8 MiB) while using less than half the power of the K80 in a relatively small die. This larger memory helps increase the operational intensity of applications to let them utilize the abundant MACs even more fully.

   We found that despite a recent emphasis on CNNs in the architecture community, they constitute only about 5% of the representative NN workload for our datacenters, which suggests more attention should be paid to MLPs and LSTMs. Repeating history, it's similar to when many architects concentrated on floating-point performance when most mainstream workloads turned out to be dominated by integer operations.

   We observed that inferences per second (IPS) is more a function of the NN than of the underlying hardware, and so IPS is an even worse single performance metric for NN processors than MIPS and MFLOPS are for CPUs and GPUs.

   We also learned that inference applications have serious response-time bounds because they are often part of user facing applications, thus NN architectures need to perform well when coping with 99th-percentile latency deadlines. While the K80 may excel at training, on average it is just a little faster than Haswell at inference for our workload, perhaps because of its emphasis on throughput rather than latency; that conflicts with the strict response-time deadlines of our inference applications.

   The TPU die leverages its advantage in MACs and on-chip memory to run short programs written using the domain-specific TensorFlow framework 15 times as fast as the K80 GPU die, resulting in a performance/Watt advantage of 29 times, which is correlated with performance/total cost of ownership. Compared to the Haswell CPU die, the corresponding ratios are 29 and 83. While future CPUs and GPUs will surely run inference faster, a redesigned TPU using circa 2015 GPU memory would go two to three times as fast and boost the performance/Watt advantage of nearly 70 over the K80 and 200 over Haswell.

   In summary, the TPU succeeded because of the large—but not too large—matrix multiply unit; the substantial software-controlled on-chip memory; the ability to run whole inference models to reduce dependence on host CPU; a single-threaded, deterministic execution model that proved to be a good match to 99th-percentile response time limits; enough flexibility to match the NNs of 2017 as well as of 2013; the omission of general-purpose features that enabled a small and low power die despite the larger datapath and memory; the use of 8-bit integers by the quantized applications; and that applications were written using TensorFlow, which made it easy to port them to the TPU at high-performance rather than them having to be rewritten to run well on the very different TPU hardware.

   Order-of-magnitude differences between commercial products are rare in computer architecture, which may lead to the TPU becoming an archetype for domain-specific architectures. We expect that many will build successors that will raise the bar even higher.

## Acknowledgements

We thank the leadership of our company for recognizing the need for a TPU and for providing the resources to build, distribute, evaluate, and publish. Special thanks go to Luiz Barroso and James Laudon for helping start the project. It takes a village to design, verify, and implement the hardware and software of a system like a TPU and to manufacture, deploy, and use it at scale, which is why there are many authors. (All authors but David Patterson worked on the TPU; he joined in 2016.) The first four authors did the bulk of the evaluation in this paper, which is why they are in front, with the rest in alphabetical order. Jouppi was also the senior architect of the whole project; the reward for his yeoman's work is being the lead author.